\documentclass[useAMS,usenatbib]{mn2e}
\usepackage{graphicx}

\title[Observational Constraints on Modified Chaplygin Gas in Horava-Lifshitz Gravity ]{Observational Constraints on  Modified Chaplygin Gas in Horava-Lifshitz Gravity 
}
\author[B. C. Paul and P. Thakur]{ B. C. Paul,$^{1,3}$ \thanks{ Electronic mail : bcpaul@iucaa.ernet.in } and
  P. Thakur,$^{2,3}$ \thanks{ Electronic mail : prasenjit \textunderscore thakur1 @yahoo.co.in}
   \\
  $^1$Physics Department, North Bengal University \\ 
      Dist. : Darjeeling, Pin : 734 013, West Bengal, India 
\\
  $^2$Physics Department, Alipurduar College\\
      Dist. : Jalpaiguri, Pin : 736122, West Bengal, India
 \\
 $^3$ IUCAA Reference Centre, Physics Department \\
 North Bengal University \\}
\date{}
\begin{document}


\pagerange{\pageref{firstpage}--\pageref{lastpage}} \pubyear{2012}

\maketitle

\label{firstpage}
\begin{abstract}

We present Cosmological models with modified Chaplygin gas (MCG) in the framework of Horava-Lifshitz (HL) theory of gravity both with and without detailed balance. The equation of state (EOS) for a MCG contains three unknown parameters namely, $A$, $\alpha$, $B$. The allowed values of some of these parameters of the EOS are determined using the recent astrophysical and cosmological observational data.
Using observational data from  $H(z)-z$, BAO peak parameter, CMB shift parameter  we study cosmologies in detailed-balance and beyond detailed-balance scenario. In this paper we  take up the beyond detailed-balance scenario in totality and contribution of dark radiation in the case of  detailed-balance scenario on the parameters of the EOS.  We explore  the effect of dark radiation on the whole range the  of  effective neutrino parameter to constrain matter contributing parameter $B$ in  both the detailed-balance  and the beyond-detailed balance scenario. It has been observed that greater the dark radiation less the matter contribution in the MCG in both the scenario considered here. In order to check the validity of beyond detailed balance scenario  we plot supernovae magnitudes ($\mu$) with redshift of Union2 data and then the variation of state parameter with redshift is studied. It has been observed that beyond detailed balance scenario is equally suitable in HL gravity with MCG.

\end{abstract}


\section{Introduction}

The big-bang cosmology has become the standard model for cosmology which accommodates a beginning of the Universe at some finite past. The discovery of CMBR \citep{pw,pw1} supports such model of the universe.  However,  big-bang cosmology based on perfect fluid assumption fails to account some of the observed facts both in the early and at late universe. The standard Big-bang model is known to have several limitations; for instance, (i) Horizon problem (ii) Flatness problem (iii) the singularity problem, to name a few.  It is known that the above problems can be resolved by invoking a phase of inflation at a very early epoch. Most of these problems have, however, been resolved by invoking inflation \citep{guth,linde,al,sato} in the semiclassical theory of gravity.
On the other hand recent observations predict that our universe is passing through a phase of acceleration \citep{ag}. This phase of acceleration is believed to be a late time phase of the universe and it comes out that such a phase cannot be accommodated in the general theory of relativity with the usual matter fields in the standard model of particle physics.
Since a universe with inflation should give satisfactory explanation of what happens close to the Planck era, it is also necessary to consider a satisfactory theory which is valid near that epoch. It may be pointed out here that a quantum gravity effect becomes important at the Planck time. But a consistent theory of quantum gravity is yet to emerge. In this direction superstring theory  may be considered as one of the promising candidate of quantum theory of gravity. Cosmological models are also proposed in Loop Quantum Gravity  (LQG) \citep{bojowald} which avoids initial singularity.
However, a proper description of time evolution of quantum space-time in the LQG is not satisfactory. Several attempts have been made in the recent past to achieve a complete quantum gravitational theory (UV complete theory). Among many such attempts,  Horava-Lifshitz (henceforth, HL) theory of gravity appears to be interesting. The success of the Lifshitz theory in solid state physics motivated Horava  to propose a theory of gravity, often called Horava-Lifshitz (HL) gravity \citep{b2} which may be important to explore a viable cosmological model. 
In the ultraviolet (UV) limit, HL gravity has a Lifshitz-like anisotropic scaling as $t\rightarrow l^{z}t$ and $x^{i}\rightarrow lx^{i}$, between space and time, characterized by the dynamical critical exponent $z = 3$ and thus breaks the Lorentz invariance; while in the infra-red (IR) limit, the scale reduces to $z = 1$. So, it is expected that it may reduce to  classical general relativistic theory of gravity in the low energy limit. The  Friedmann equation gets modified by a  $\frac{1}{a^{4}}$ term \citep{b3,calcagni,kiritsis}, where $a$ is the scale factor in a non-flat universe in the HL-gravity.

In the original HL gravity, Horava assumed two conditions: detailed balance and projectibility. More recently, Sotiriou, Visser and Weifurtner (SVW) \citep{sotiriou}, proposed  a general HL theory with projectability but without detailed-balance conditions. For a spatially curved Friedmann-Robertson-Walker universe, the SVW generalization yields  an extra  $\frac{1}{a^{6}}$ -term that modifies the coefficient of the  $\frac{1}{a^{4}}$ term in the Friedmann equation  as compared to the HL theory. Therefore, it is important to look  for cosmological models in Horava gravity considering projectibility  with and without detailed-balance.

In the HL-gravity, the  initial bigbang singularity may not arise due to the presence of higher order terms in the spatial curvatures $R_{ij}$ [\ref{s1}]. There are many such novel features of HL gravity for which it is worth to explore different aspects of observed universe.
A volume of literature in the framework of the HL gravity appeared containing the study of gravitational wave production \citep{mukohyama,park1,myung1}, perturbation spectrum \citep{gao,cai1,wang1}, black-hole properties \citep{danielson,cai2,kehagias}, dark energy phenomenology
\citep{park,chaichian}, the problems of determining observational constraints in the theory \citep{dutta}, astrophysical phenomenology \citep{kim,harko,iorio}, thermodynamical properties etc. \citep{wang2,cai3}. Though there exists foundational and  conceptual issues of Horava-Lifshitz gravity and its associated cosmology, cosmological scenario have been examined with generalised Chaplygin gas (GCG) \citep{ali}. GCG being an exotic matter may be useful to address the recent acceleration of the universe. One of the characteristic features of the GCG is that it behaves as a pressureless fluid at the early stage of the evolution of the universe, and at a later stage it behaves as a  cosmological constant.
Recently, a modified form of Chaplygin gas is also considered extensively in cosmology
\citep{liu,thakur}. The modified Chaplygin gas (MCG) is more general and contains three free parameters. The idea is to interpolate states of standard fluids at high pressures and at high energy densities to a constant negative pressure at low energy densities \citep{debnath}. 
In the present work we explore cosmological models with MCG in the framework of HL gravity and determine the range of parameters of MCG from recent cosmological observations. We examine the effect of effective neutrino parameter on both detailed-balance and beyond-detailed balance scenario. Here compatibility of  beyond detailed balance scenario  is also explored in details using recent observational data in HL gravity with MCG . The objective of the paper is to determine the limits of the unknown EOS parameters using the observational data. The equation of state parameter of the total cosmic fluid defined by  $w(z)=\frac{p_{tot}}{\rho_{tot}}$, will be evaluated  and examined at  different values of redshift parameters. Comparing the supernovae magnitudes  ($\mu$) vs. redshift ($z$) with Union2 data we test the viability of beyond-detailed balance scenario. The suitability of the model is also examined using $w(z)$ vs. $z$ plot in our model.

The paper is organized as follows: In sec. 2, we present the basic equations for Horava-Lifshitz cosmology and obtain  the Friedmann equations for detailed balance and beyond detailed balance conditions. In sec. 3, the energy density and EOS for MCG is presented. In sec. 4 , the  constraints on detailed-balance  condition and beyond detailed balance condition from the observations is presented. In sec. 5, numerical analysis to  determine constraints on EOS parameters are obtained for detailed balance. In sec. 6, numerical analysis to  determine constraints on EOS parameters are obtained for beyond detailed balance scenario. In sec. 7, the viability of MCG in HL gravity is discussed. Finally, in sec. 8, we  summarize the result.

\section{ Horava-Lifshitz Cosmology}

In Horava-Lifshitz gravity \citep{calcagni,kiritsis}, it is convenient to use the four-dimensional space-time metric of the Arnowitt-Deser-Misner (ADM) decomposition form which is given by
\begin {equation}
ds^{2}=-N^{2}dt^{2}+g_{ij}(dx^{i}+N^{i}dt)(dx^{j}+N^{j}dt)
\end{equation}
where the basic variables are lapse function $N$, shift vector $N_{i}$, and the spatial metric $g_{ij}$. The scaling transformation of the co-ordinates reads: $t\rightarrow l^{3}t$ and $x^{i}\rightarrow lx^{i}$. The shift $N^{i}$ and the 3d spatial metric  $g_{ij}$ depend both on the time coordinate $t$ and the spatial coordinate $x^{i}$, the lapse   $N$ is assumed to depend on time only. This condition on the lapse is called the projectibility condition.
The action of HL gravity consists of kinetic and potential terms as follows:
\begin {equation}
S_{g}=S_{K}+S_{V}=\int dt \; d^{3}x\sqrt{g} \; N(L_{K}+L_{V}).
\end {equation}
The kinetic term is given by
\begin{eqnarray}
S_{K}=\int dt d^{3}x\sqrt{g}N \left[\frac{2(K_{ij}K^{ij}-\lambda K^2)}{\kappa^2}\right]
\end{eqnarray} 
where $
K_{ij}=\frac{(\dot {g_{ij}}-\nabla_{i}N_{j}-\nabla_{j}N_{i})}{2N} $
is the extrinsic curvature and dot represents derivative with respect to time ($t$).

\subsection{Detailed Balance Condition and Projectibility}
The symmetry property of the Lagrangian $L_{V}$, employed in  the gravitational action drastically reduces the number of invariants which one should actually consider in the action to begin with \citep{horava2}. The above symmetry is known as detailed balance which follows from   condensed matter systems and requires that the Lagrangian $L_{V}$ should be derivable from a superpotential $W$ \citep{kiritsis}.
Under the detailed balance condition the total action of HL gravity is given by

\begin{eqnarray}
\label{s1}
S_{g} = \int dt d^{3}x\sqrt{g}N \left[\frac{2(K_{ij}K^{ij}-\lambda K^2)}{\kappa^2}+\frac{\kappa^2 C_{ij}C^{ij}}{2\omega^4} \right. \nonumber \\       -\frac{\kappa^2\mu\epsilon^{ijk} R_{il}\nabla_{j}R_{k}^{l}}{2\omega^2\sqrt{g}}+\frac{\kappa^2\mu^2 R_{ij}R^{ij}}{8} \nonumber \\
 \left[ -\frac{k^2\mu^2}{8(3\lambda-1)}\left[\frac{(1-4\lambda)R^{2}}{4}+\Lambda R-3\Lambda^{2}\right]\right]
\end{eqnarray} 
where
\begin{equation}
C^{ij}=\frac{\epsilon^{ijk}}{\sqrt{g}} \nabla_{k} \left(R^{j}_{i}-\frac{R\delta^{j}_{i}}{4} \right)
\end{equation}
is known as the Cotton tensor, and the covariant derivatives are determined with respect to the spatial metric ($g_{ij}$),     $\epsilon^{ijk}$ is a totally antisymmetric unit tensor, $\lambda$ is a dimensionless constant and the variables $\kappa$, $\omega$ and $\mu$ are constants .

In the above gravitational action to include  matter components one needs to add a cosmological stress-energy tensor to the gravitational field equations, that recovers the usual general relativity formulation in the low-energy limit \citep{sotiriou,chaichian,carloni}. The matter-tensor is a hydrodynamical approximation that leads to the existence of energy density ($\rho_{m}$) and pressure ($p_{m}$) in the Friedmann equation, where $\rho_{m}$ represents the total matter energy density, that accounts for both the baryonic $\rho_{b}$ as well as that of the dark matter $\rho_{dm}$,  including the normal matter (where $p_{m}$ represents pressure). 

Horava obtained the gravitational action assuming that the lapse function is just a function of time i.e.,  $N = N(t)$. 
Here we use FRW metric with  
$N = 1$, $g_{ij}=a^{2}(t)\gamma_{ij}$, $N^{i}=0$
with  
\begin{equation}
\gamma_{ij}dx^{i}dx^{j}=\frac{dr^{2}}{1-Kr^2}+r^2d\Omega^{2}_{2}
\end{equation}
where $K=-1,1,0$, corresponds to open, close and flat universe respectively. By varying $N$ and $g_{ij}$ in the gravitational action (4), one obtains the following field equations:
\begin{eqnarray}
H^{2} &=& \frac{\kappa^{2}}{6(3\lambda-1)}(\rho_{m}+\rho_{r})  \nonumber \\ 
 &&+\frac{\kappa^{2}}{6(3\lambda-1)} \left[\frac{3\kappa^{2}\mu^{2}K^{2}}{8(3\lambda-1)a^4} +\frac{3\kappa^{2}\mu^{2}\Lambda^{2}}{8(3\lambda-1)}\right]     \nonumber  \\
&& -\frac{\kappa^{4}\mu^{2}\Lambda K}{8(3\lambda-1)^2a^2} 
\end{eqnarray}

\begin{eqnarray}
\dot{H}+\frac{3H^{2}}{2}&=&-\frac{\kappa^{2}}{4(3\lambda-1)}(\rho_{m}\omega_{m}+\rho_{r}\omega_{r})  \nonumber \\ &&-\frac{\kappa^{2}}{4(3\lambda-1)}\left[\frac{\kappa^{2}\mu^{2}K^{2}}{8(3\lambda-1)a^4}- 
\frac{3\kappa^{2}\mu^{2}\Lambda^{2}}{8(3\lambda-1)}\right] \nonumber \\
&& -\frac{\kappa^{4}\mu^{2}\Lambda K}{16(3\lambda-1)^2a^{2}} 
\end{eqnarray}
where $H=\frac{\dot{a}}{a}$.
In the above field equations the term proportional to $a^{-4}$ may be considered as the usual $"dark \, radiation \,term"$, characteristics of the HL Cosmology \citep{calcagni,kiritsis} and the constant term  is identified  with the usual cosmological constant.
The conservation equation for matter is:  
\begin{equation}
\dot{\rho_{m}}+3H(\rho_{m}+p_{m})=0,
\end{equation}
and that of  radiation is: 
\begin{equation}
\dot{\rho_{r}}+3H(\rho_{r}+p_{r})=0,
\end{equation}
where we denote
\begin{equation}
G_{cosmo}=\frac{\kappa^{2}}{16\pi(3\lambda-1)},
\end{equation}
\begin{equation}
\frac{\kappa^{4}\mu^{2}\Lambda}{8(3\lambda-1)^{2}}=1,
\end{equation}
\begin{equation}
 G_{grav}=\frac{\kappa^{2}}{32\pi}.
\end{equation}

\subsection{Beyond Detailed Balance Condition with Projectibility}

As it is not known with certainty whether the detailed balance condition is enough for extracting  whole information of Horava-Lifshitz gravity \citep{calcagni,kiritsis}
 or it is necessary to do something with this balance condition, one can investigate cosmological scenario in  the HL gravity relaxing 
the detailed balance condition. In this subsection we discuss  the cosmology of HL gravity in the pressence of modified Chaplygin gas (MCG), baryon, radiation and dark radiation without detailed-balance. The aim of the paper is to look for the effects of the dark-radiation on the parameters of the MCG model. The Friedmann equations in this case can be written as 
\citep{sotiriou1,carloni,bogdanos,charmousis,leon}:
\begin{eqnarray}
H^{2} &=& \frac{2\sigma_{0}}{(3\lambda-1)}(\rho_{m}+\rho_{r})  \nonumber \\ 
 &&+\frac{2}{(3\lambda-1)} \left[\frac{\sigma_{1}}{6}+\frac{\sigma_{3}K^{2}}{6a^4} +\frac{\sigma_{4}K}{6a^6}\right] \nonumber \\  
 &&+\frac{\sigma_{2} K}{3(3\lambda-1)a^2},
\end{eqnarray}

\begin{eqnarray}
\dot{H}+\frac{3}{2}H^{2}&=& \frac{3\sigma_{0}}{(3\lambda-1)}(\rho_{m}\omega_{m}+\rho_{r}\omega_{r})  \nonumber \\ 
 &&-\frac{3}{(3\lambda-1)} \left[-\frac{\sigma_{1}}{6}+\frac{\sigma_{3}K^{2}}{18a^4} +\frac{\sigma_{4}K}{6a^6}\right] \nonumber \\
&&+\frac{\sigma_{2} K}{6(3\lambda-1)a^2},
\end{eqnarray} 
where $\sigma_{0}=\kappa^2/12$.

The dimensionless parameters are given by:
\begin{equation}
G_{cosmo}=\frac{6\sigma_{0}}{8\pi(3\lambda-1)},
\end{equation}
\begin{equation}
\sigma_{2}=-3(3\lambda-1),
\end{equation}
\begin{equation}
G_{grav}=\frac{6\sigma_{0}}{16\pi},
\end{equation}
where  $\sigma_{2} < 0$  and $\sigma_{4} > 0$.
In the case of the detailed balance case, in the IR limit ($\lambda=1$), the two parameters $G_{cosmo}$ and $G_{grav}$ coincides.

\section{EOS for Modified Chaplygin Gas}

The equation of state for Generalized Chaplygin Gas (GCG) \citep{billic,bento} is given by
\begin{equation}
p=-\frac{A}{\rho^\alpha}
\end{equation}
with $ 0 \leq \alpha \leq 1 $. In the above  original Chaplygin gas corresponds to $\alpha=1$ \citep{chap}.
It may be pointed out here that
Chaplygin introduced his equation of state \citep{chap}  to study the lifting force on a plane wing in aerodynamics.
Chaplygin's equation of state (19) has raised recently a renewed interest \citep{ch1,ch2}  because of its many remarkable and, in some sense, intriguingly unique features. It is interesting to note that it has amazing connection with string theory: it can be obtained from the  Nambu-Goto action for d-branes moving in a (d+2)-dimensional spacetime in the light-cone parametrization \citep{ch3}.
However, in cosmological context generalized form of Chaplygin gas  may be useful to describe the observed universe. GCG  has two free parameters $A$ (positive), $\alpha$. 
Recently a further modification of GCG has been proposed in the framework of cosmology \citep{liu}. The modified Chaplygin gas (MCG) is more general and it contains one more free parameter ($B$). The model is consistent with (i) Gravitational lensing test  \citep{silva,dev} and (ii) Gamma-ray bursts \citep{bertolami}.
The equation of state for the MCG is given by:
\begin{equation}
\label{mcgeos}
p=B\rho-\frac{A}{\rho^\alpha}
\end{equation}
where $A$, $B$, $\alpha$ are arbitrary constants to be determined from observation for model building 
with $0 \leq \alpha \leq 1 $.

The energy conservation equation for the MCG is:
\begin{equation}
\label{mcgconsv}
\dot{\rho_{c}}+3H(\rho_{c}+p_{c})=0
\end{equation}
where $\rho_{c}$ and $p_{c}$ correspond to energy density and pressure of MCG respectively.
Using eq. (\ref{mcgeos}) in eq.(\ref{mcgconsv})  we obtain:
\begin{equation}
\label{rhoc1}
\rho_{c} = \left[\frac{A}{1+B} + \frac{C}{a^{3n}} \right]^{\frac{1}{1+ \alpha}}
\end{equation}
where $C$ is an arbitrary  constant and we denote $(1 + B)(1+ \alpha) = n $.
Equation (\ref{rhoc1}) can be rewritten as 
\begin{equation}
\rho_{c}=\rho_{o}\left[A_{S}+\frac{1-A_{S}}{a^{3n}}\right]^{\frac{1}{1+
\alpha}}
\end{equation}
where we denote $ A_{S} = \frac{A}{1+B}\frac{1}{\rho_{o}^{\alpha +1}}$ with $\frac{a}{a_{o}}=\frac{1}{1+z}$, $z$ is redshift parameter and we choose $a_{o}=1$ for convenience. MCG reduces to GCG model when we set $B=0$ in the above equation.

\section{Observational Constraints on  EOS Parameters}

In general theory of relativity cosmological models with MCG has been studied and the constraints on EOS parameters for viable cosmologies are determined using observational data \citep{thakur, lu}. The EOS parameters of MCG will be explored here for viable cosmologies in the  framework of  Horava-Lifshitz gravity  using the recent observational data. For this we have taken up data from Observed Hubble Data (OHD), BAO peak parameter and CMB shift parameter.

\subsection{Constraints Obtained from Detailed Balance}

In this case, using eqs. (7) and (8), the Friedmann equations can be rewritten as:
\begin{eqnarray}
H^{2} &=& \frac{8\pi G}{3} \left(\rho_{b}+\rho_{c}+\rho_{r} \right)+\left(\frac{K^{2}}{2\Lambda  a^{4}} 
 +\frac{\Lambda}{2}\right) \nonumber \\ 
&&-\frac{K}{a^2},
\end{eqnarray}

\begin{eqnarray}
\dot{H}+\frac{3}{2}H^{2} &=& -4\pi G \; (p_{c}+\frac{1}{3}\rho_{r})-\left( \frac{K^{2}}{4\Lambda a^{4}}
          -\frac{3\Lambda}{4}\right) \nonumber \\ 
&&-\frac{K}{2a^{2}}.
\end{eqnarray}

Let us define the following dimensionless density parameters:

(i) for matter component
\begin{equation}
\hspace{2 cm} \Omega_{i}\equiv\frac{8 \pi G}{3H^{2}} \; \rho_{i}
\end{equation}

(ii) for curvature
\begin{equation}
\hspace{2 cm} \Omega_{K}\equiv -\frac{K}{H^{2}a^{2}}
\end{equation}

(iii) for cosmological constant 
\begin{equation}
\hspace{2 cm} \Omega_{o}\equiv \frac{\Lambda}{2H^{2}_{o}}.
\end{equation}

We define another  dimensionless parameter for  expansion rate as:
\begin{equation}
\hspace{2 cm} E(z)\equiv\frac{H(z)}{H_{o}}.
\end{equation}
Using the above definition of parameters, the Friedmann equation now can be rewritten as:
\begin{eqnarray}
E^{2}(z)&=&\Omega_{bo}(1+z)^{3}+\Omega_{co}F(z) +\Omega_{ro}(1+z)^{4} \nonumber \\ 
&& +\Omega_{Ko}(1+z)^{2} + \left( \Omega_{o}+\frac{\Omega^{2}_{Ko}(1+z)^{4}}{4\Omega_o} \right), 
\end{eqnarray}
where
\begin{equation}
F(z)=\left[A_{S}+\frac{1-A_{S}}{a^{3(1+B)(1+\alpha)}}\right]^{\frac{1}{1+
\alpha}}.
\end{equation}
Let us assume  $E(z=0)=1$ at the present epoch, which leads to
\begin{equation}
\label{energy}
\Omega_{bo}+\Omega_{co}+\Omega_{ro}+\Omega_{Ko}
+
\Omega_{o}+\frac{\Omega^{2}_{Ko}}{4\Omega_o}=1
\end{equation}
where $\Omega_{bo}$, $\Omega_{co}$, $\Omega_{ro}$, $\Omega_{Ko}$ represent the present day baryon, MCG, radiation and curvature energy density parameters respectively. Here $\Omega_{o}$ is the energy density associated with the cosmological constant.
The last term in eq. (\ref{energy}) corresponds to dark radiation, which is a characteristic feature of the Horava-Lifshitz theory of gravity. The dark radiation component may be important during  nucleosynthesis. Thus a suitable bound from Big Bang Nucleosynthesis 
(henceforth, BBN) may be incorporated in the above EOS. Using the upper limit on the total amount of Horava-Lifshitz dark radiation that is permitted  during BBN era is expressed by the parameter $\Delta N_{\nu}$ which represents the effective neutrino species \citep{hagiwara,olive}. We  obtain the following constraint equation \citep{dutta}:
\begin{equation}
\label{radiation}
\hspace{2 cm} \frac{\Omega^{2}_{Ko}}{4\Omega_{o}}=0.135\Delta N_{\nu}\Omega_{ro}
\end{equation}
The BBN upper limit on $\Delta N_{\nu}$ is  $-1.7\leq \Delta N_{\nu}\leq 2.0$, is taken from the {\it Refs. } \citep{olive,steigman}. A negative value of $\Delta N_{\nu}$  is
usually associated with models involving decay of a massive particles which we do not consider here. Again $\Delta N_{\nu} = 0$, which corresponds to the zero curvature scenario will be  excluded also. It is because of the fact that the Horava-Lifshitz cosmology with zero curvature becomes indistinguishable from $\Lambda CDM$. The curvature in dynamical dark energy models are important, neglecting the curvature term impose a serious problem \citep{clarkson,virey}. Therefore, we consider the limiting values for $\Delta N_{\nu}$ which satisfies the bound  $0< \Delta N_{\nu}\leq 2.0$.

The numerical analysis taken up here contains nine  parameters, these are namely, 
$\Omega_{bo}$, $\Omega_{co}$, $\Omega_{ro}$, $\Omega_{Ko}$, $\Omega_{o}$, $\Delta N_{\nu}$,    
$H_{o}$, $A_{S}$, $B$, $\alpha$.  As the number of unknowns are more than the number of  equations we  fix some of the parameters using the best-fit values from 7 year WMAP data \citep{komatsu}. The fixed parameters are $\Omega_{mo}(\equiv\Omega_{bo}+\Omega_{co})$, $\Omega_{bo}$, $H_{o}$, $\Omega_{ro}$ and the corresponding values of the parameters are chosen as follows : $\Omega_{mo}=0.27$, $\Omega_{bo}=0.04$, $H_{o}=71.4Km/sec/Mpc$, $\Omega_{ro}=8.14*10^{-5}$.
Therefore,  one can have now only six free parameters to be determined which are $\Omega_{Ko}$, $\Omega_{o}$, $A_{S}$, $B $, $\alpha$, $\Delta N_{\nu}$.
Using eq. (\ref{radiation})  in eq. (\ref{energy}) one obtains
\begin{eqnarray}
\Omega_{o}(K,\Delta N_{\nu},A_{S},\alpha)=1-\Omega_{mo}-(1-0.135\Delta N_{\nu})\Omega_{ro}\nonumber \\-0.73 (K)\sqrt{\Delta N_{\nu}}\sqrt{\Omega_{ro}-\Omega_{mo}\Omega_{ro}-\Omega^{2}_{ro}}
\end{eqnarray}
\begin{equation}
\Omega_{Ko}(\Delta N_{\nu},A_{S},\alpha)=\sqrt{0.54 \Delta N_{\nu} \Omega_{ro} \Omega_{o}(K, \Delta N_{\nu}, A_{S}, \alpha)}
\end{equation}
which may be employed for a close or in an open universe depending on the values of K.
Now, it reduces to four free parameters, namely,  $A_{S}$, $B$, $\alpha$, $ \Delta N_{\nu}$.
To determine the effect of dark radiation on the constraints of the parameters of the MCG in detailed-balance scenario, we took two extreme values of $\alpha$ ($\alpha$=$0.999$, $0.001$) satisfying $0 \leq \alpha \leq1$  for two extreme values of $\Delta N_{\nu}$ (0.01,  2.0) in both close and open universe  respectively. In this case each of these values of $\alpha$ and $\Delta N_{\nu}$ determined the best-fit values of the rest two parameters ({\it i.e.,} $A_{S}$, $B$). Thereafter at the extreme values of $\Delta N_{\nu}$ for two extreme values of $\alpha$ we plot contours for the parameters  $A_{S}$, $B$ at different confidence levels. From the contours of  $A_{S}$, $B$ drawn at different values of $\alpha$ and  $\Delta N_{\nu}$ we determine  the permissible range of values of the $B$-parameter for the MCG in HL gravity in the framework of open or closed universe.

\subsection{Constraints Obtained from Beyond-Detailed Balance}

In beyond-detailed balance scenario using eqs. (14)-(15), the Friedmann's equations can be rewritten as:
\begin{eqnarray}
H^{2} &=& \frac{8\pi G}{3}(\rho_{b}+\rho_{c}+\rho_{r})  \nonumber \\ 
 &&+ \left[\frac{\sigma_{1}}{6}+\frac{\sigma_{3}K^{2}}{6a^4} +\frac{\sigma_{4}K}{6a^6}\right] \nonumber \\  
 &&-\frac{K}{a^2} 
\end{eqnarray}

\begin{eqnarray}
\dot{H}+\frac{3}{2}H^{2}&=& -4\pi G(p_{c}+\frac{1}{3}\rho_{r}) \nonumber \\ 
 &&-\frac{3}{2} \left[-\frac{\sigma_{1}}{6}+\frac{\sigma_{3}K^{2}}{18a^4} +\frac{\sigma_{4}K}{6a^6}\right] \nonumber \\   
&&-\frac{K}{2a^2} 
\end{eqnarray}

Now one can re-write the dimensionless Hubble parameter as follows:

\begin{eqnarray}
E^{2}(z)&=&\Omega_{bo}(1+z)^{3}+\Omega_{co}F(z) +\Omega_{ro}(1+z)^{4} + \Omega_{Ko}(1+z)^{2} \nonumber \\ 
&&+ [\Omega_{1}+\Omega_{3}(1+z)^{4}+\Omega_{4}(1+z)^{6}] 
\end{eqnarray}
 where
\begin{equation}
F(z)=\left[A_{S}+\frac{1-A_{S}}{a^{3(1+B)(1+\alpha)}}\right]^{\frac{1}{1+
\alpha}}.
\end{equation}
The dimensionless parameters namely,  $\Omega_{1}$, $\Omega_{3}$, $\Omega_{4}$ are related to the model parameters $\sigma_{1}$, $\sigma_{3}$, $\sigma_{4}$ as follows:

\begin{equation}
\Omega_{1}=\frac{\sigma_{1}}{6H^2_{o}},
\end{equation}

\begin{equation}
\Omega_{3}=\frac{\sigma_{3}H^2_{o}\Omega^2_{Ko}}{6},
\end{equation}

\begin{equation}
\Omega_{4}=-\frac{\sigma_{4}\Omega_{Ko}}{6}.
\end{equation}
At the present epoch $E(z=0)=1$, which leads to
\begin{equation}
\label{energybdb}
\Omega_{bo}+\Omega_{co}+\Omega_{ro}+\Omega_{Ko}
+
\Omega_{1}+\Omega_{3}+\Omega_{4}=1
\end{equation}

In the above equations $\Omega_{4}$ is required to be a positive quantity in order that the Hubble parameter remains positive at all redshifts, also the gravitational perturbations \citep{sotiriou1,bogdanos} demand the same. For the conveneince of our analysis  $\Omega_{3}$ is assumed positive definite.

The above equation is a constraint equation for this analysis and we use it to replace $\Omega_{1}$ in terms of other parameters in our analysis.
Following the procedure adopted in Ref. \citep{dutta}  for $\Delta N_{\nu}$, we consider the upper limit of dark radiation  beyond standard model at the BBN. Consequently, the following constraints at the time of BBN emerged ($z=z_{BBN}$) \citep{hagiwara,olive,steigman,malaney} :
\begin{equation}
\label{radiationbdb}
\Omega_{3}+\Omega_{4}(1+z^2_{BBN})^2=\Omega_{3max}=0.135\Delta N_{\nu}\Omega_{r0}.
\end{equation}
where the $\Omega_{3}$ represent the usual dark radiation and $\Omega_{4}$ represents a kinetic-like component (a quintessence field dominated by kinetic energy) \citep{joyce1,joyce2}.
The above equation will be used to replace $\Omega_{4}$ in terms of other parameters in the analysis.
For simplicity we define 
\begin{equation}
\beta=\frac{\Omega_{3}}{\Omega_{3max}}
\end{equation}
where $\Omega_{3max}$ is the upper limit on $\Omega_{3}$.
This will help us to express  $\Omega_{3}$ in terms of other parameters.

Following the detailed-balance scenario we consider $\Delta N_{\nu}$ here to satisfy the bound  $0<\Delta N_{\nu}\leq 2.0$, following the importance of curvature in dark energy models and treating $\Omega_{Ko}$ as a free parameter \citep{clarkson,virey}.

To sum up, in the numerical analysis taken up here, the following parameters,
$\Omega_{bo}$, $\Omega_{co}$, $\Omega_{ro}$, $\Omega_{Ko}$, $\Omega_{1}$, $\Omega_{3}$, $\Omega_{4}$, $\Delta N_{\nu}$,    
$H_{o}$, $A_{S}$, $B$, $\alpha$, $\beta$ are involved. We fix some of the parameters using the best-fit values from 7 year WMAP data \citep{komatsu}. The fixed parameters are $\Omega_{mo}(\equiv\Omega_{bo}+\Omega_{co})$, $\Omega_{bo}$, $H_{o}$, $\Omega_{ro}$ and the corresponding values of the parameters are chosen as follows : $\Omega_{mo}=0.27$, $\Omega_{bo}=0.04$, $H_{o}=71.4Km/sec/Mpc$, $\Omega_{ro}=8.14*10^{-5}$. 
Using the constraint eqs. (36) - (42) one can  replace $\Omega_{1}$, $\Omega_{3}$, $\Omega_{4}$ in terms of the other six free parameters for the numerical analysis.
Therefore one can have only six free parameters to be determined which are $\Omega_{Ko}$, $A_{S}$, $B$, $\alpha$, $\beta$, $\Delta N_{\nu}$.

To determine the constraints on the parameters of the MCG in beyond detailed-balace scenario, we consider three values of $\alpha$ satisfying $0 \leq \alpha \leq1$ ($\alpha$=$0.999$, $0.500$, $0.001$)  and determine the best-fit values for the rest five parameters (i.e., $A_{S}$, $B$, $\beta$, $\Omega_{Ko}$, $\Delta N_{\nu}$). Thereafter, at the best-fit values of $\Delta N_{\nu}$, $\beta$, $\Omega_{Ko}$ for those three values of $\alpha$ we plot 2d contours  for the parameters  $A_{S}$, $B$  at different confidence levels. The contours of  $A_{S}$, $B$ drawn at different values of $\alpha$ in turn determine  the permissible range of values of the $B$-parameter for the MCG in HL gravity in the framework of  beyond detailed balance scenario.

To examine the effect of dark radiation (i.e., effective neutrino parameter) on the constraints on the parameters of the MCG, we took two extreme values of $\alpha$ ($\alpha$=$0.999$, $0.001$) satisfying $0 \leq \alpha \leq1$  for two extreme values of $\Delta N_{\nu}$ (0.01, 2.0). In this case each of these values of $\alpha$ , $\Delta N_{\nu}$ determines the best-fit values of the rest four parameters (i.e., $A_{S}$, $B$, $\beta$, $\Omega_{Ko}$). Thereafter, at the extreme values of $\Delta N_{\nu}$ for two extreme values of $\alpha$ we plot 2d contours  for the parameters  $A_{S}$, $B$ for the best-fitted values of $\beta$, $\Omega_{Ko}$ at different confidence levels. From the contours of  $A_{S}$, $B$ drawn at different values of $\alpha$ and  $\Delta N_{\nu}$ we determine  the permissible range of values of the $B$-parameter for the MCG in HL gravity in the framework of  beyond detailed balance scenario. We note that the range of values of $B$ is narrower due to the effect of  effective neutrino parameter on  $B$.

\section{Numerical Analysis to determine constraints on the EOS parameters in detailed balance scenario}

In this section we use three sets of different observational data  to constrain the parameters of the MCG. Stern data set for (H-z) data has been used along with BAO peak parameter and CMB shift parameter.
Chi-square minimisation technique has been used here to determine the limiting values of the EOS parameters in the next subsections.

\subsection{(H-z) data as a tool for constraining}

The best-fitted parameters of the model considered here can be obtained by minimising the entity chi-square which is defined as
\begin{eqnarray}
\chi^{2}_{OHD}(H_{o},A_{S}, B ,\alpha,\Delta N_{\nu},z) =  \\
 \sum\frac{(H(H_{o}, A_{S}, B,\alpha, \Delta N_{\nu},z)-H_{obs}(z))^2}{\sigma^{2}_{z}}
\end{eqnarray}
where $H_{obs}(z)$ is the observed Hubble parameter at redshift(z) and $\sigma^{2}_{z}$ is the associated error with that particular observation.
The Hubble parameter is given by
\begin{equation}
H(z)=H_{o} E(z)
\end{equation} 
where
\begin{eqnarray}
E(z)&=& (\Omega_{bo}(1+z)^3+\Omega_{co}F(z)+\Omega_{ro}(1+z)^4 \nonumber \\
&&+\Omega_{Ko}(1+z)^2+  \left(\Omega_{o}+\frac{\Omega^{2}_{Ko}(1+z)^4}{4\Omega_o} \right) )^{1/2} 
\end{eqnarray}
and denoting
\begin{equation}
F(z)=\left[A_{S}+\frac{1-A_{S}}{a^{3(1+B)(1+\alpha)}}\right]^{\frac{1}{1+
\alpha}}.
\end{equation}
Here $H(z)-z$ data is taken from Stern Data analysis \citep{stern}.
There are 12 data points of $H(z)$ at redshift $z$ used to constrain the MCG model.

\subsection{BAO peak parameter as a tool for constraining}

 A model independent BAO (Baryon Acoustic Oscillation) peak parameter can be defined for low redshift ($z_{1}$) measurements as:
\begin{equation}
\label{baop}
\mathcal{A}=\frac{\sqrt{\Omega_{m}}}{\left[E\left(z_{1}\right)\right]^{\frac{1}{3}}}\left[\frac{\int_{0}^{z_{1}}\frac{dz}{E\left(z\right)}}{z_{1}}\right]^{\frac{2}{3}}
\end{equation}
where $\Omega_{m}$ is the matter density parameter for the Universe. For a detailed description of the above defined parameter and related approximations reader is referred to \citep{bao}. The chi square function can be defined as:
\begin{equation}
\label{chib}
\chi^{2}_{BAO}=\frac{\left(\mathcal{A}-0.469\right)^{2}}{ \left(0.017\right)^{2}}
\end{equation}
where we have used the measured value for $\mathcal{A}$ ($0.469\pm.0.017$) as obtained by \citep{bao} from the SDSS data for LRG (Luminous Red Galaxies) survey.

\subsection{CMB Shift Parameter as a tool for constraining}

Here the CMB shift parameter is defined as
\begin{equation}
R=\sqrt{\Omega_{m}}\int_{0}^{z_{ls}}\frac{dz}{E(z)}
\end{equation}
where $z_{ls}$ is the z at last scattering.
The WMAP7 data gives us 
$R=1.726\pm0.018$ at $z=1091.3$ \citep{komatsu}.
Chi square in this case is defined as 
\begin{equation}
\chi^{2}_{CMB}=\frac{(R-1.726)^{2}}{(0.018)^{2}}
\end{equation}

\subsection{Joint Analysis with (H-z)+BAO+CMB }

Total chi-square function for our joint analysis:
\begin{equation}
\label{chitbd}
\chi^{2}_{tot}=\chi^{2}_{OHD}+\chi^{2}_{BAO}+\chi^{2}_{CMB}
\end{equation}
The statistical analysis with $\chi^{2}_{tot}$ gives the bounds on the model parameter specially on $B$.

\begin{figure}
\centering
{\subfigure{[a]\includegraphics[width=230pt,height=190pt]{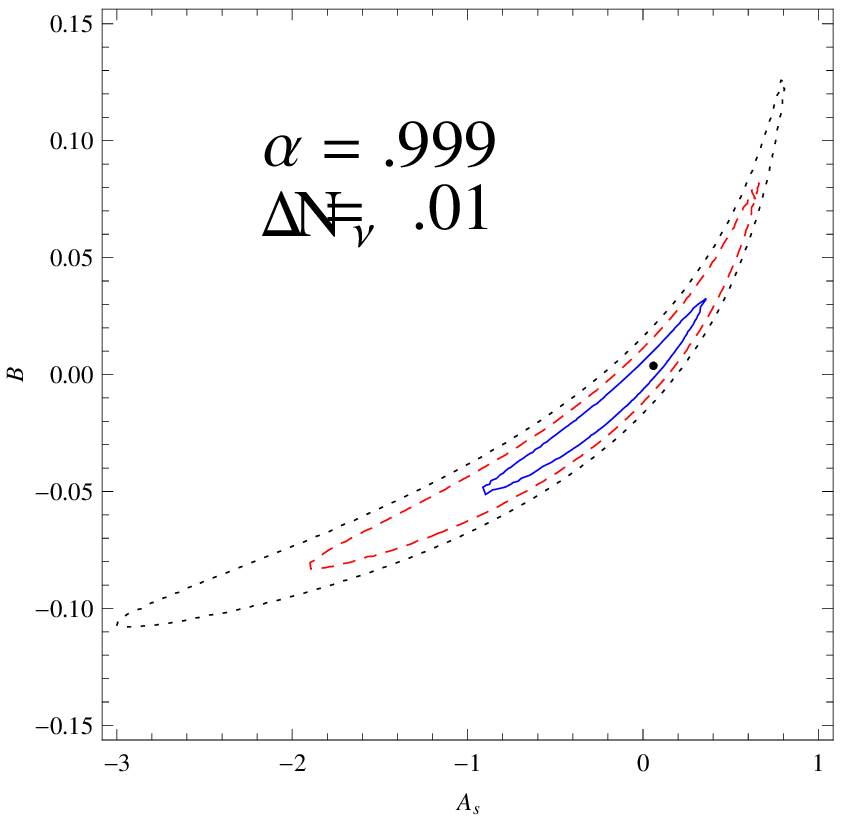}} \\
\subfigure{[b]\includegraphics[width=230pt,height=190pt]{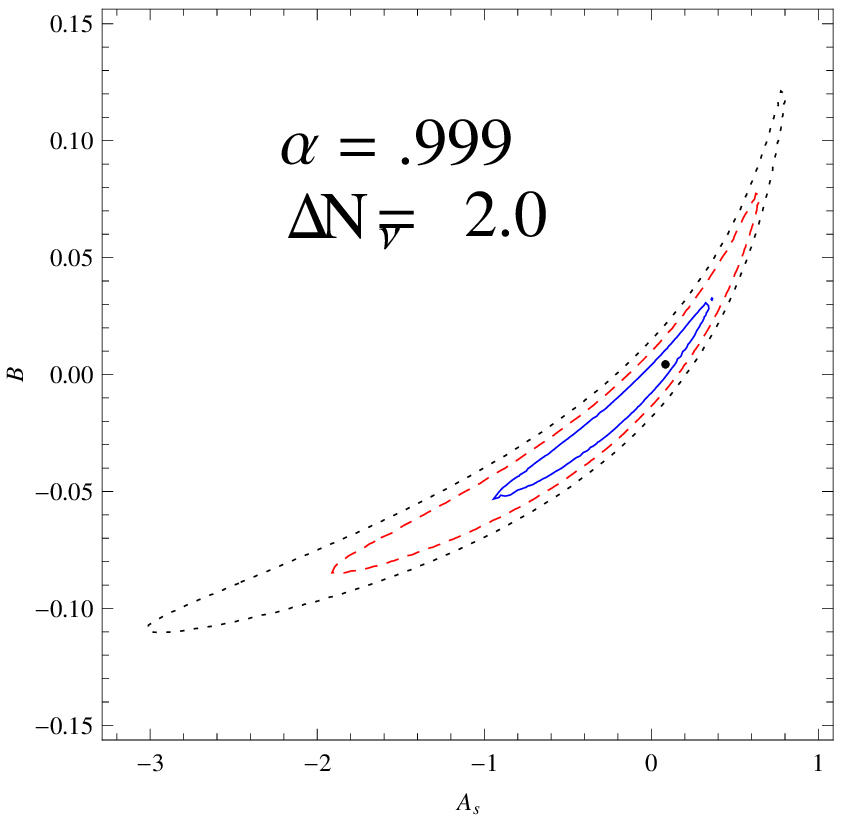}}} \\
\caption{Constraints for closed universe for $\alpha=0.999$ from OHD+SDSS+CMB Shift data    $68.3\%$(Solid) $95.4\%$ (Dashed) and $99.73 \%$  (Dotted) contours.}
\label{DB-a}
\end{figure}

\begin{figure}
\centering
{\subfigure{[a]\includegraphics[width=230pt,height=190pt]{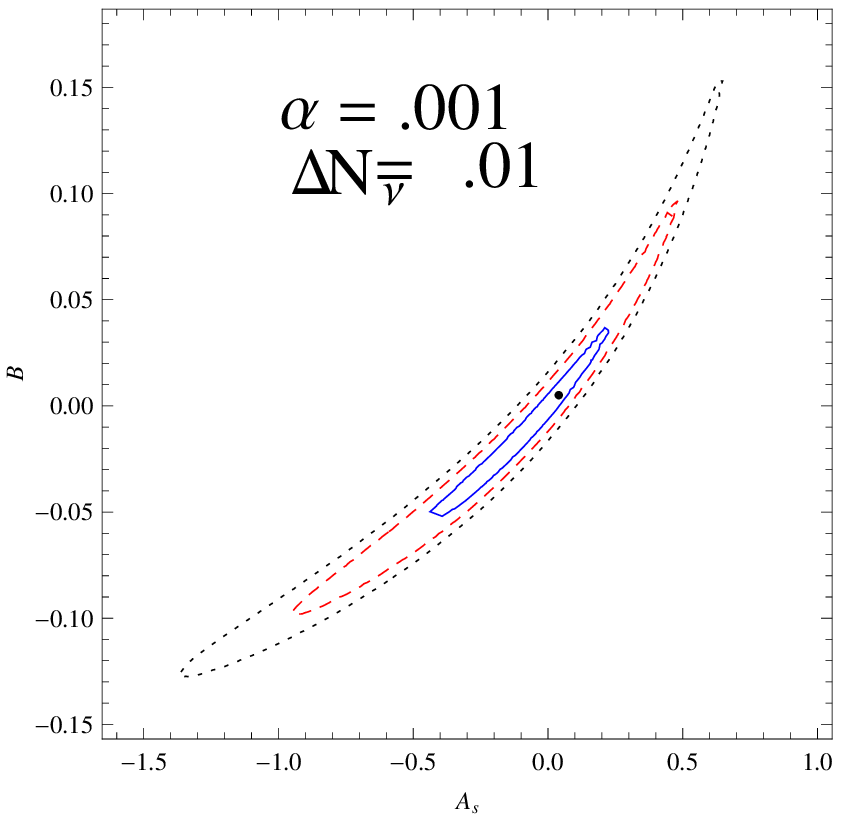}} \\
\subfigure{[b]\includegraphics[width=230pt,height=190pt]{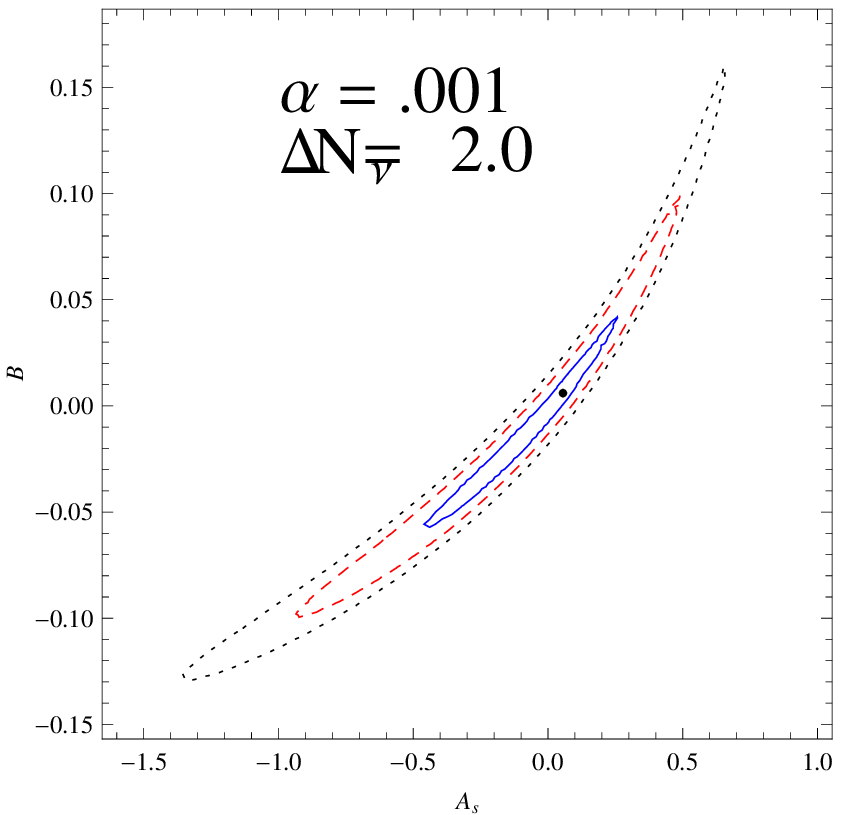}}} \\
\caption{Constraints for closed universe for $\alpha=0.001$ from OHD+SDSS+CMB Shift data    $68.3\%$(Solid) $95.4\%$ (Dashed) and $99.73 \%$  (Dotted) contours.}
\label{DB-b}
\end{figure}

\begin{table}
\centering
		\begin{tabular}{@{}lccr}
		\hline
		Model     & $B$ &    $A_{S}$ &    $\Delta N_{\nu}$\\
		\hline
		$\alpha=0.999$ & 0.003745 & 0.062817 &  0.232994\\
		$\alpha=0.500$ & 0.016592 & 0.110548 &  0.099996\\
		$\alpha=0.001$ & 0.006192 & 0.052076 &  0.807051\\
		\hline	
		\end{tabular}
\caption{Best-fit values for MCG: $K=1$}	
	\label{thub1}
\end{table}

\begin{table}
\centering
		\begin{tabular}{@{}lccr}
		\hline
		Model                                  & $B$ &      $A_{S}$    \\
		\hline
		$\alpha=0.999$,$\Delta N_{\nu}=0.01$ & 0.00374357 & 0.0593243 \\
		$\alpha=0.999$,$\Delta N_{\nu}=2.00$ & 0.00434955 & 0.0841938 \\
		$\alpha=0.001$,$\Delta N_{\nu}=0.01$ & 0.00504535 & 0.0400150 \\
		$\alpha=0.001$,$\Delta N_{\nu}=2.00$ & 0.00600186 & 0.0555941 \\
		\hline	
		\end{tabular}
\caption{Best-fit values for MCG: $K=1$}	
	\label{thub1a}
\end{table}

The contours between $B$ and $A_{S}$ for closed universe for $\alpha=0.999$ and $\alpha=0.001$ are shown in  figs. (1) and (2) respectively. Figures 1(a), 2(a) are drawn for $\Delta N_{\nu}$=0.01 and Figures 1(b) and 2(b) are drawn for $\Delta N_{\nu}$=2.0.
From fig. 1(a) which is plotted for $\alpha=0.999$ and $\Delta N_{\nu}$=.01 for closed universe it appears that the value of $B$ lies in the range $-0.05138 <B<0.03351$, $-0.08234 <B<0.08444$, $-0.1073 <B<0.1274$  at $68.3\%$, $95.4\%$, $99.73\%$ confidence levels respectively. 
From fig. 1(b) which is plotted for $\alpha=0.999$ and $\Delta N_{\nu}$=2.0 for closed universe, the value of $B$ lies in the range $-0.05338 <B<0.03251$, $-0.08434 <B<0.07745$, $-0.1103 <B<0.1224$  at $68.3\%$, $95.4\%$, $99.73\%$ confidence levels respectively.
It is evident from the analysis that the range of values of $B$ decreases with an increase of the effective neutrino parameter $\Delta N_{\nu}$.

From fig. 2(a) which is plotted for $\alpha=0.001$ and $\Delta N_{\nu}$= 0.01 for closed universe, the value of $B$ lies in the range $-0.05042 <B<0.03751$, $-0.09878 <B<0.09906$, $-0.1274 <B<0.1562$  at $68.3\%$, $95.4\%$, $99.73\%$ confidence levels respectively. 
From fig. 2(b) which is plotted for $\alpha=0.001$ and $\Delta N_{\nu}$=2.0 for closed universe, the value of $B$ lies in the range $-0.05702 <B<0.0419$, $-0.09988 <B<0.1002$, $-0.1296 <B<0.1595$  at $68.3\%$, $95.4\%$, $99.73\%$ confidence levels respectively.
It is clear from the above analysis that the range of values of $B$  increases with an increase in the effective neutrino parameter $\Delta N_{\nu}$. Thus for a given vlaue of $\alpha$, the increase in neutrino parameter $\Delta N_{\nu}$ increases the range of $B$ parameter in the positive side.

\begin{figure}
\centering
{\subfigure{[a]\includegraphics[width=230pt,height=190pt]{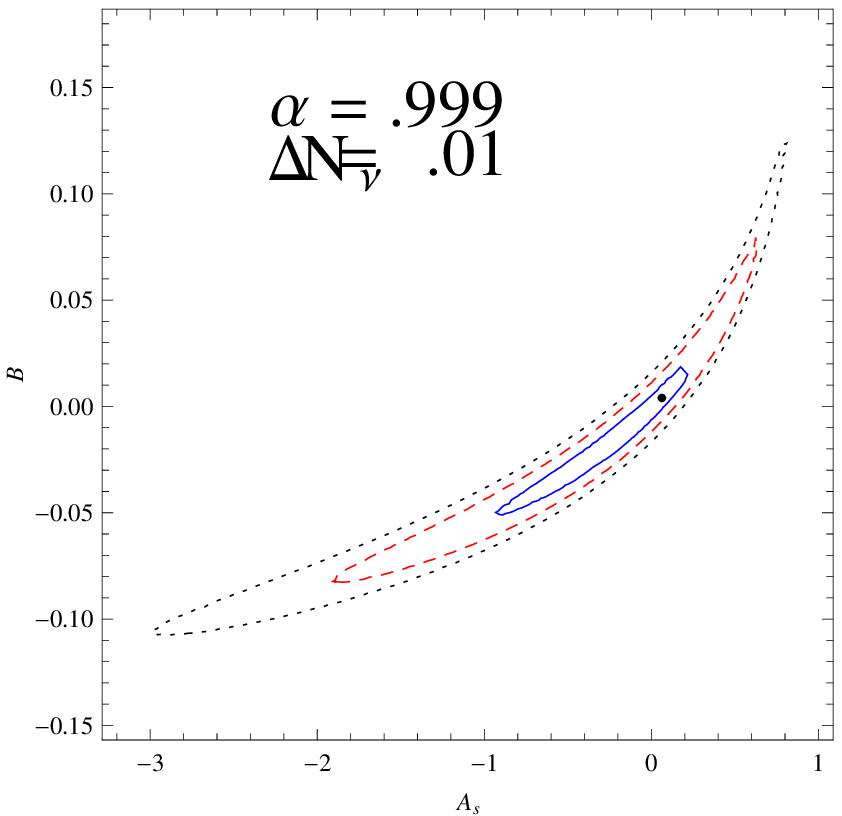}} \\
\subfigure{[b]\includegraphics[width=230pt,height=190pt]{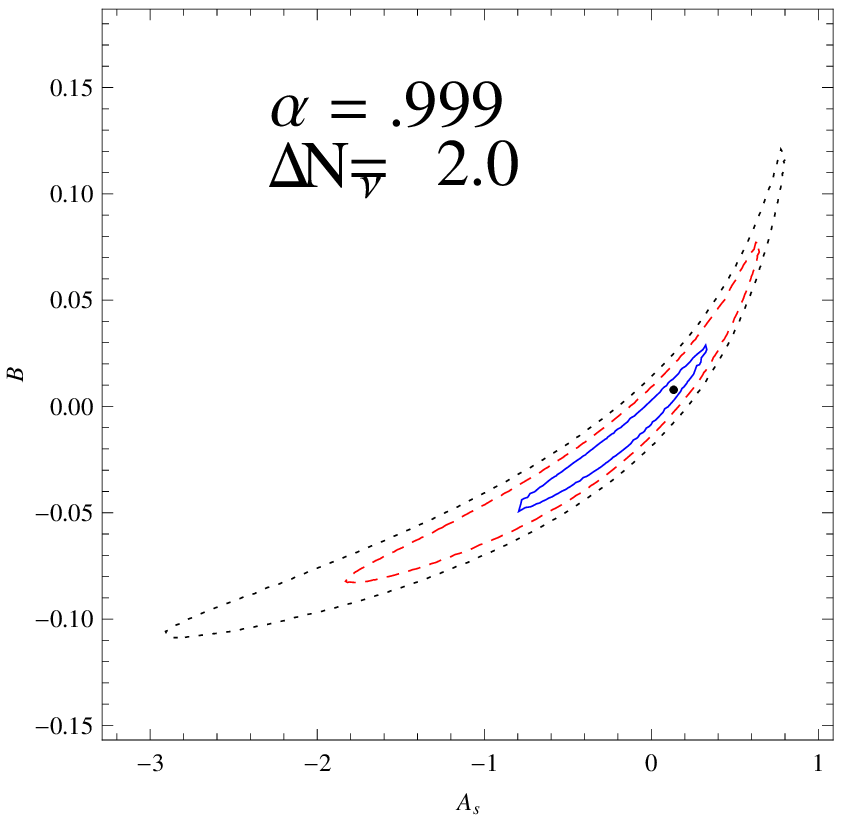}}} \\
\caption{Constraints for open universe for $\alpha=0.999$ from OHD+SDSS+CMB Shift data    $68.3\%$(Solid) $95.4\%$ (Dashed) and $99.73 \%$  (Dotted) contours.}
\label{DB-c}
\end{figure}

\begin{figure}
\centering
{\subfigure{[a]\includegraphics[width=230pt,height=190pt]{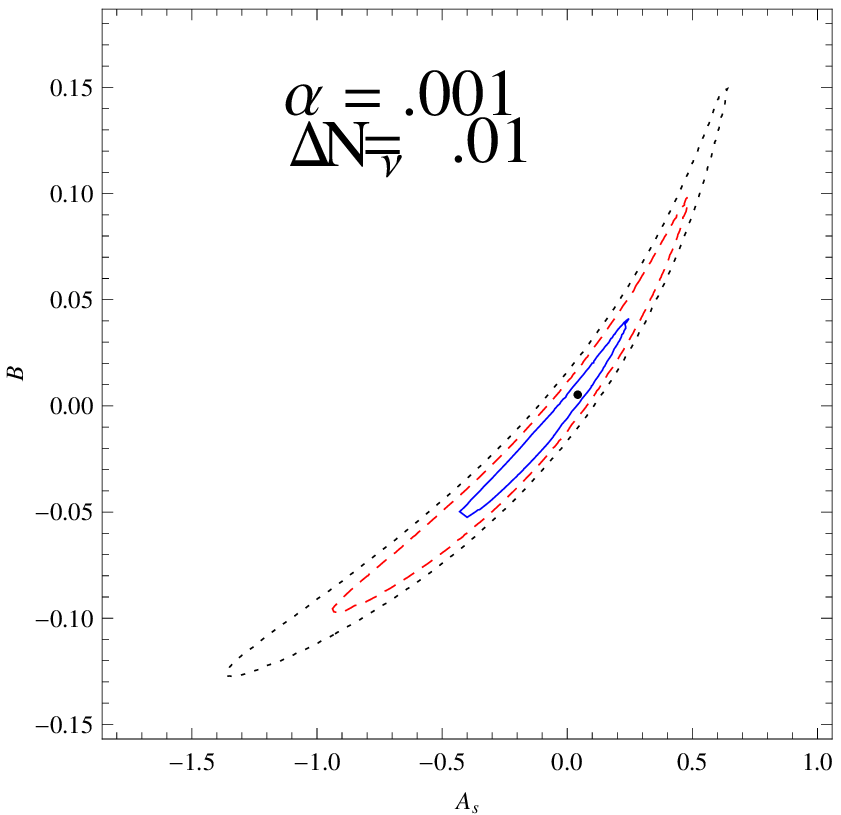}} \\
\subfigure{[b]\includegraphics[width=230pt,height=190pt]{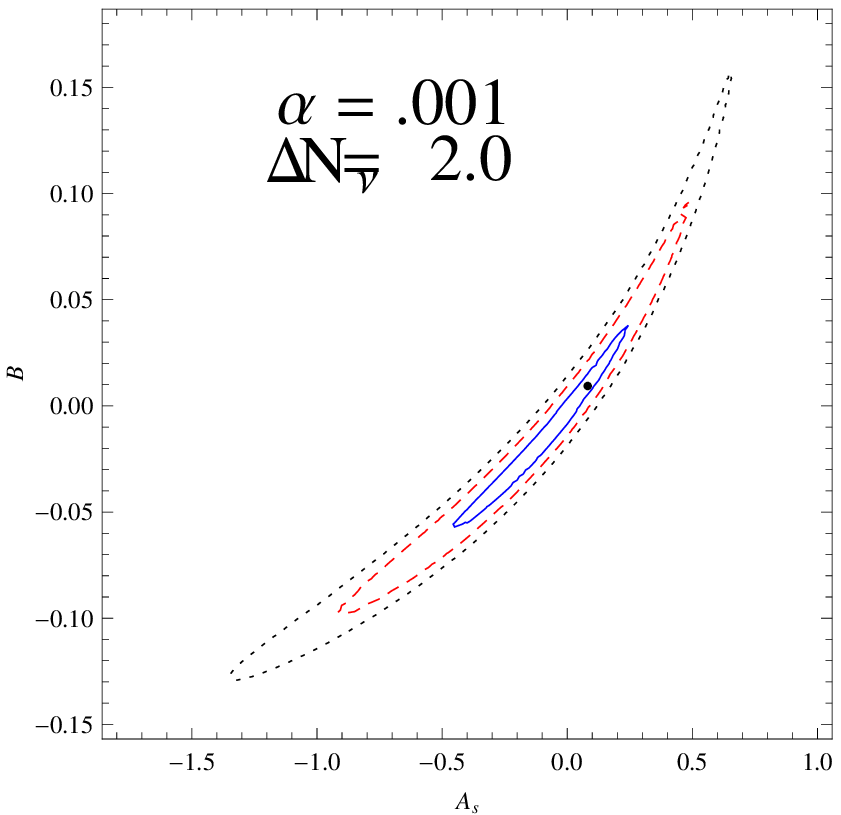}}} \\
\caption{Constraints for open universe for $\alpha=0.001$ from OHD+SDSS+CMB Shift data    $68.3\%$(Solid) $95.4\%$ (Dashed) and $99.73 \%$  (Dotted) contours.}
\label{DB-d}
\end{figure}

\begin{table}
\centering
		\begin{tabular}{@{}lccr}
		\hline
		Model     & $B$ &    $A_{S}$ &    $\Delta N_{\nu}$  \\
		\hline
		$\alpha=0.999$ & 0.007498 & 0.107866 &  0.100055 \\
		$\alpha=0.500$ & 0.010499 & 0.110576 &  0.100002 \\
		$\alpha=0.001$ & 0.016478 & 0.114298 &  0.100005\\
		\hline	
		\end{tabular}
\caption{Best-fit values for MCG : $K=-1$}	
	\label{thub3}
\end{table}
\begin{table}
\centering
		\begin{tabular}{@{}lccr}
		\hline
		Model                                  & $B$ &      $A_{S}$    \\
		\hline
		$\alpha=0.999$,$\Delta N_{\nu}=0.01$ & 0.00398596 & 0.0629336 \\
		$\alpha=0.999$,$\Delta N_{\nu}=2.00$ & 0.00782675 & 0.133407 \\
		$\alpha=0.001$,$\Delta N_{\nu}=0.01$ & 0.00528027 & 0.0418902 \\
		$\alpha=0.001$,$\Delta N_{\nu}=2.00$ & 0.00937448 & 0.0817538 \\
		\hline	
		\end{tabular}
\caption{Best-fit values for MCG: $K=-1$}	
	\label{thub3a}
\end{table}

The contours between $B$ and $A_{S}$ for open universe for $\alpha=0.999$ and $\alpha=0.001$  are drawn in figs. (3) and (4) respectively.  Figures 3 (a) and 4 (a)  are drawn for $\Delta N_{\nu}$ =0.01 and Figures 3 (b) and 4(b)  are drawn for $\Delta N_{\nu}$=2.0.
It is evident from fig. 3(a) which is plotted for $\alpha=0.999$ and $\Delta N_{\nu}$= 0.01 for open universe,  that the parameter $B$ satisfies the following inequalities  $-0.05141<B<0.0178$, $-0.08437 <B<0.07932$, $-0.1063 <B<0.1266$  at $68.3\%$, $95.4\%$, $99.73\%$ confidence level respectively. 
In the fig. 3 (b), the allowed range of values of the parameter $B$ for  $\alpha=0.999$ and $\Delta N_{\nu}$=2.0 for open universe are obtained which are given by $-0.04921 <B<0.02879$, $-0.08437 <B<0.07713$, $-0.1074<B<0.120$  at $68.3\%$, $95.4\%$, $99.73\%$ confidence levels  respectively.
It is evident that the domain of $B$ decreases with an increase of the effective neutrino parameter $\Delta N_{\nu}$.

Figure 4(a)  is plotted for $\alpha=0.001$ and $\Delta N_{\nu}$=.01 for open universe. In this case  $B$ lies in the following ranges $-0.05262 <B<0.03971$, $-0.09768 <B<0.09686$, $-0.1274 <B<0.1496$  at $68.3\%$ , $95.4\%$ ,$99.73\%$ confidence levels respectively. 
From fig. 4(b) which is plotted for $\alpha=0.001$ and $\Delta N_{\nu}$=2.0 for open universe, one can obtain viable cosmologies where $B$ satisfies the following inequalities   $-0.05812 <B<0.03751$, $-0.09768 <B<0.09576$, $-0.1274 <B<0.1584$  at $68.3\%$ , $95.4\%$ , $99.73\%$ confidence levels respectively.
It is evident from the contours drawn in fig. (4)  that the range of $ B $ increases with an increase in the effective neutrino parameter $\Delta N_{\nu}$. But the positive range of values of $B$ decreases.

\section{Numerical Analysis to determine constraints on the EOS parameters in beyond-detailed balance scenario}

In this section we use data  to constrain the parameters of the MCG that were used in detailed balance scenario of the previous section. Stern data set for ($H-z$) data has been used along with BAO peak parameter and CMB shift parameter.
Chi-square minimisation technique has been used to determine the limiting values of the EOS parameters in the next subsections.

\subsection{(H-z) data as a tool for constraining}

The best-fitted parameters of the model considered here can be obtained by minimising the entity chi-square which is defined as

\begin{eqnarray}
\chi^{2}_{OHD}(H_{o},\Omega_{Ko},A_{S}, B ,\alpha,\beta,\Delta N_{\nu},z) =  \\
 \sum\frac{(H(H_{o},\Omega_{Ko},A_{S},B ,\alpha,\beta, \Delta N_{\nu},z)-H_{obs}(z))^2}{\sigma^{2}_{z}}
\end{eqnarray}
where $H_{obs}(z)$ is the observed Hubble parameter at redshift ($z$) and $\sigma^{2}_{z}$ is the associated error with that particular observation.
Hubble parameter is given by
\begin{equation}
H(z)=H_{o} E(z)
\end{equation} 
where we denote
\begin{eqnarray}
E^{2}(z)&=& \Omega_{bo(1+z)^{3}+\Omega_{co}F(z) +\Omega_{ro}(1+z)^{4}  +\Omega_{Ko}(1+z)^{2} \nonumber \\
&&+ [\Omega_{1}+\Omega_{3}(1+z)^{4}+\Omega_{4}(1+z)^{6}] 
\end{eqnarray}
 with 
\begin{equation}
F(z)=\left[A_{S}+\frac{1-A_{S}}{a^{3(1+B)(1+\alpha)}}\right]^{\frac{1}{1+
\alpha}}.
\end{equation}
In this case $H(z)-z$ data is taken from Stern Data analysis \citep{stern}.
There are 12 data points of $H(z)$ at redshift $z$ used to constrain the MCG model.

\subsection{BAO peak parameter as a tool for constraining}

 A model independent BAO (Baryon Acoustic Oscillation) peak parameter can be defined for low redshift ($z_{1}$) measurements as:
\begin{equation}
\label{baopbdb}
\mathcal{A}=\frac{\sqrt{\Omega_{m}}}{\left[E\left(z_{1}\right)\right]^{\frac{1}{3}}}\left[\frac{\int_{0}^{z_{1}}\frac{dz}{E\left(z\right)}}{z_{1}}\right]^{\frac{2}{3}}
\end{equation}
where $\Omega_{m}$ is the matter density parameter for the Universe. For a detailed description of the above defined parameter and related approximations reader is referred to \citep{bao}. The chi-square function can be defined as usual:
\begin{equation}
\label{chibbdb}
\chi^{2}_{BAO}=\frac{\left(\mathcal{A}-0.469\right)^{2}}{ \left(0.017\right)^{2}}
\end{equation}
where we have used the measured value for $\mathcal{A}$ ($0.469\pm.0.017$) as obtained by \citep{bao} from the SDSS data for LRG (Luminous Red Galaxies) survey.

\subsection{CMB Shift Parameter as a tool for constraining}

Here the CMB shift parameter is defined as
\begin{equation}
R=\sqrt{\Omega_{m}}\int_{0}^{z_{ls}}\frac{dz}{E(z)}
\end{equation}
where $z_{ls}$ is the z at last scattering.
The WMAP7 data gives us 
$R=1.726\pm0.018$ at $z=1091.3$ \citep{komatsu}.
Chi square is defined as 
\begin{equation}
\chi^{2}_{CMB}=\frac{(R-1.726)^{2}}{(0.018)^{2}}
\end{equation}

\subsection{Joint Analysis with (H-z) +BAO+CMB}

We define total chi-square function for our joint analysis as:
\begin{equation}
\label{chitbdb}
\chi^{2}_{tot}=\chi^{2}_{OHD}+\chi^{2}_{BAO}+\chi^{2}_{CMB}.
\end{equation}
The statistical analysis with $\chi^{2}_{tot}$ gives the bounds on the model parameter specially on $B$.
\begin{figure}
\centering
{\subfigure{[a]\includegraphics[width=230pt,height=190pt]{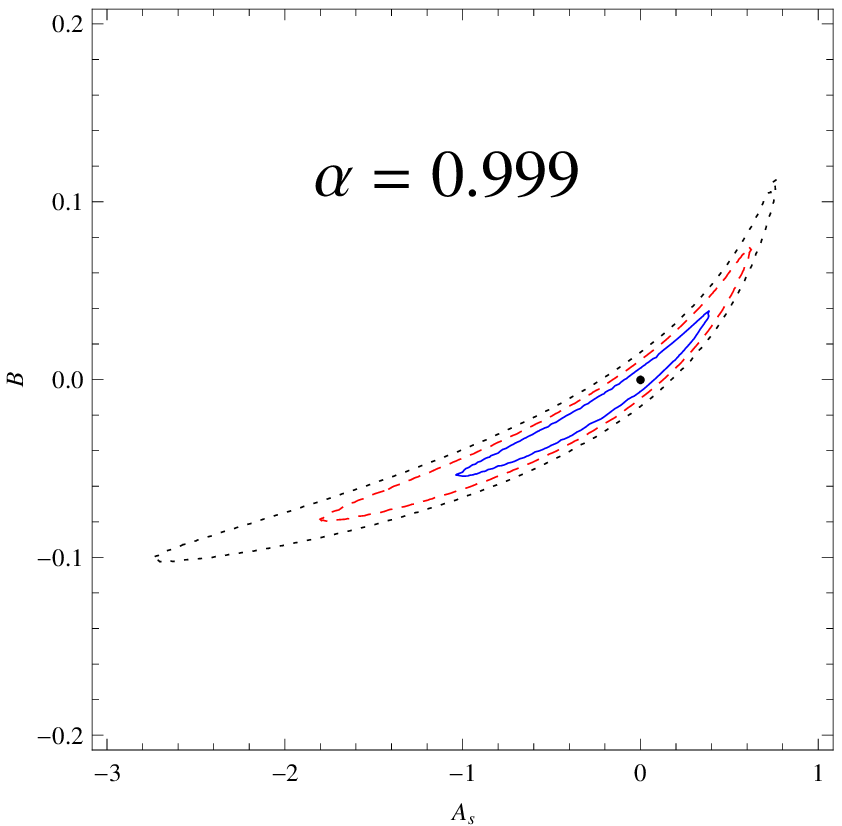}}\\
\subfigure{[b]\includegraphics[width=230pt,height=190pt]{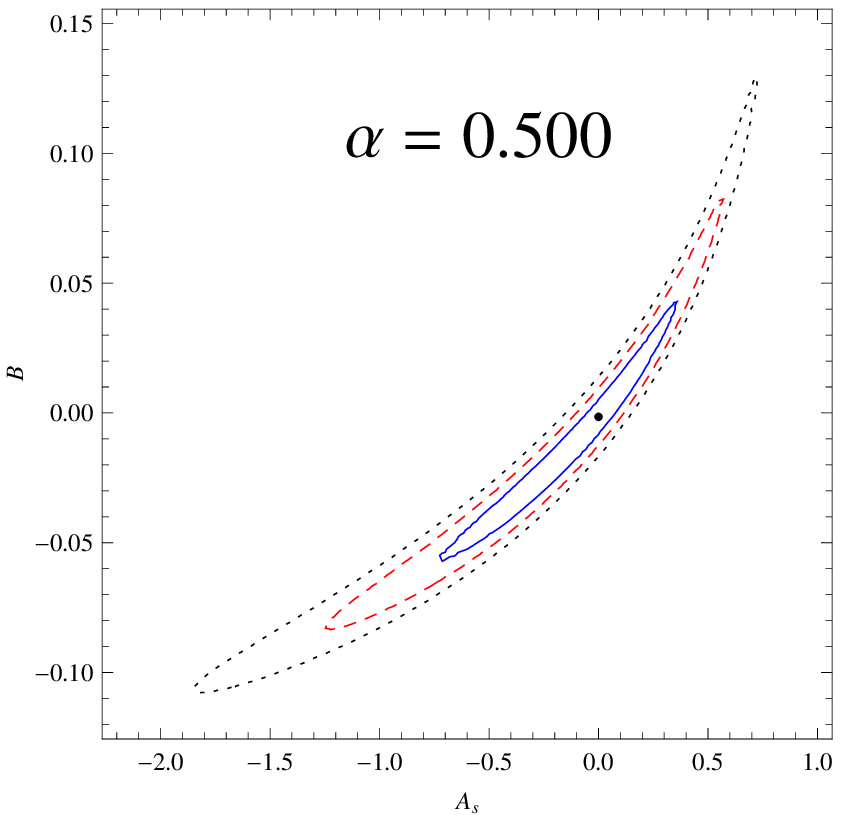}}\\
\subfigure{[c]\includegraphics[width=230pt,height=190pt]{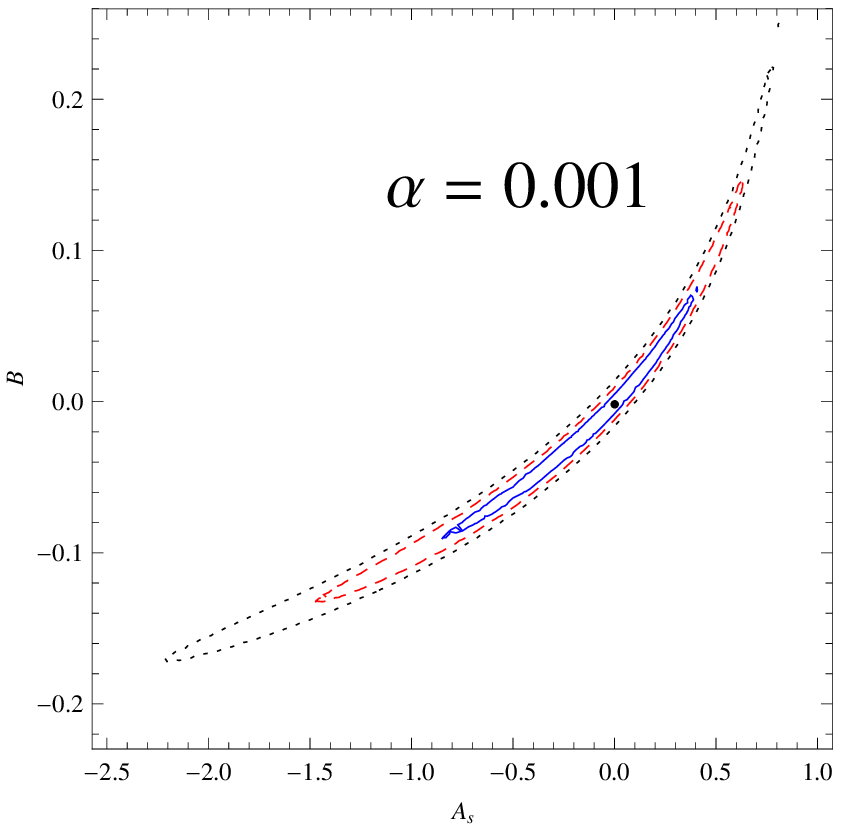}}}\\
\caption{Constraints in beyond detailed balance for $\alpha=0.999$, $\alpha=0.500$,$\alpha=0.001$,from OHD+SDSS+CMB Shift data    $68.3\%$(Solid) $95.4\%$ (Dashed) and $99.73 \%$  (Dotted) contours.}
\label{BDB-a}
\end{figure}

Fig. 5(a) is plotted for $\alpha=0.999$  with best-fitted values of $\beta$, $\Delta N_{\nu}$  and $\Omega_{Ko}$. The parameter $B$ 
satisfies the following inequalities $-0.05498 <B<0.03808$, $-0.07989 <B<0.07346$, $-0.1035 <B<0.1128$  at $68.3\%$, $95.4\%$, $99.73\%$ confidence levels respectively. 
Fig. 5(b)  is plotted for $\alpha=0.500$ for best-fitted values of $\beta$, $\Delta N_{\nu}$  and $\Omega_{Ko}$. The parameter $B$
in this case satisfies the folowing inequalities: $-0.05741 <B<0.04421$, $-0.08349 <B<0.08287$, $-0.1060<B<0.1305$  at $68.3\%$ , $95.4\%$, $99.73\%$ confidence levels respectively.  Fig. 5(c)  is plotted for $\alpha=0.001$ for best-fitted value of $\beta$, $\Delta N_{\nu}$  and $\Omega_{Ko}$. We note that the parameter $B$ satisfies the following inequalities $-0.09257 <B<0.0707$ , $-0.1326 <B<0.1493$, $-0.1727<B<0.2247$  at $68.3\%$, $95.4\%$, $99.73\%$ confidence levels respectively.
It is evident that the allowed range of values of the parameter $B$  becomes larger compared to that of the detailed balance scenario \citep{paulprd}.

\begin{figure}
\centering
{\subfigure{[a]\includegraphics[width=230pt,height=190pt]{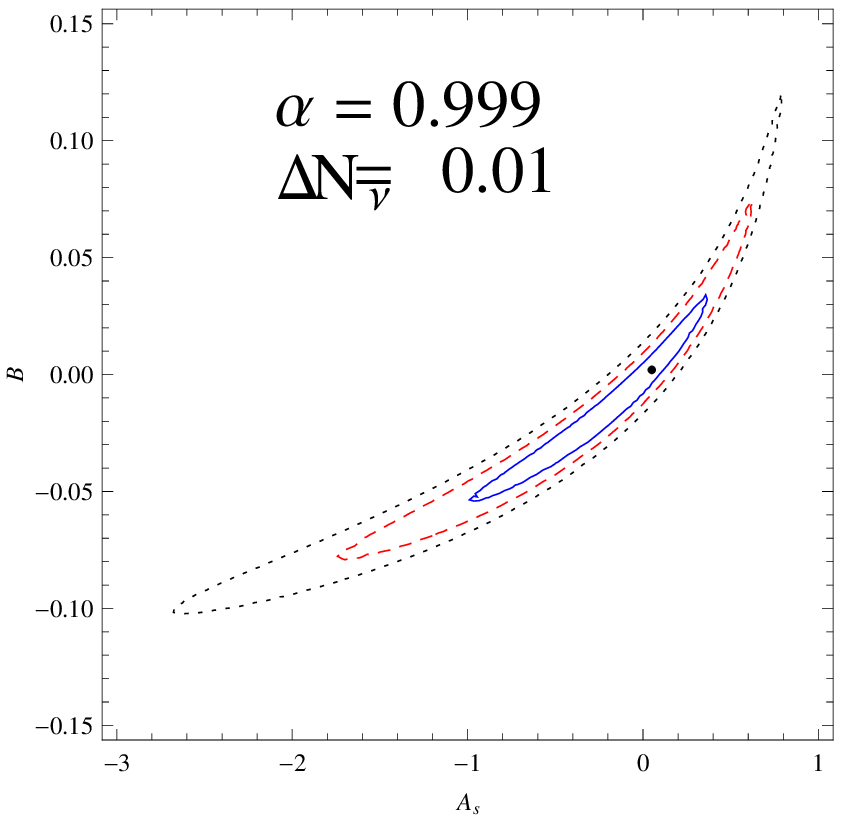}} \\
\subfigure{[b]\includegraphics[width=230pt,height=190pt]{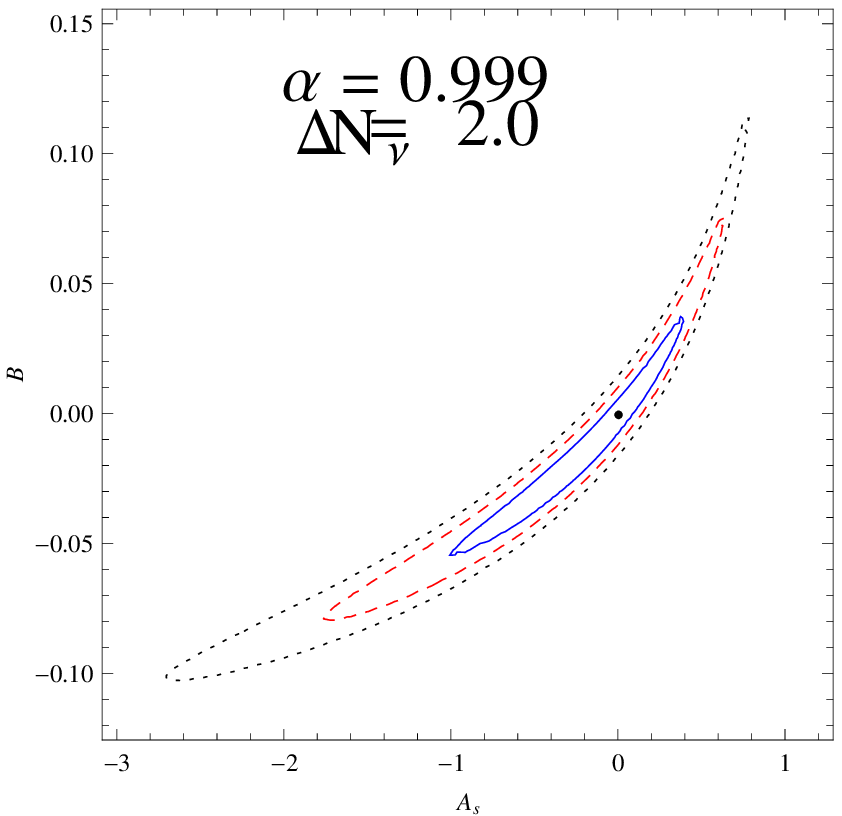}}} \\
\caption{Constraints in beyond-detailed balance for $\alpha=0.999$ from OHD+SDSS+CMB Shift data    $68.3\%$(Solid) $95.4\%$ (Dashed) and $99.73 \%$  (Dotted) contours.}
\label{BDB-c}
\end{figure}

Fig. 6(a) is plotted for $\alpha=0.999$ and $\Delta N_{\nu}$=0.01 for best-fitted value of $\beta$ and $\Omega_{Ko}$, it is evident that $B$ can take any value in the following ranges $-0.05338 <B<0.03351$, $-0.07935 <B<0.07445$, $-0.1013 <B<0.1224$  at $68.3\%$,  $95.4\%$, $99.73\%$ confidence levels respectively. 
Fig. 6(b)  is plotted for $\alpha=0.999$ and $\Delta N_{\nu}$=2.0, it is evident that the value of B lies in the range $-0.05462 <B<0.03617$, $-0.07978 <B<0.07662$, $-0.1032<B<0.1162$  at $68.3\%$, $95.4\%$,  $99.73\%$ confidence levels respectively.
The figs. 6(a)- 6(b) show that the range of permissible values of  $B$ decreases with an increase in the effective neutrino parameter.

\begin{figure}
\centering
{\subfigure{[a]\includegraphics[width=230pt,height=190pt]{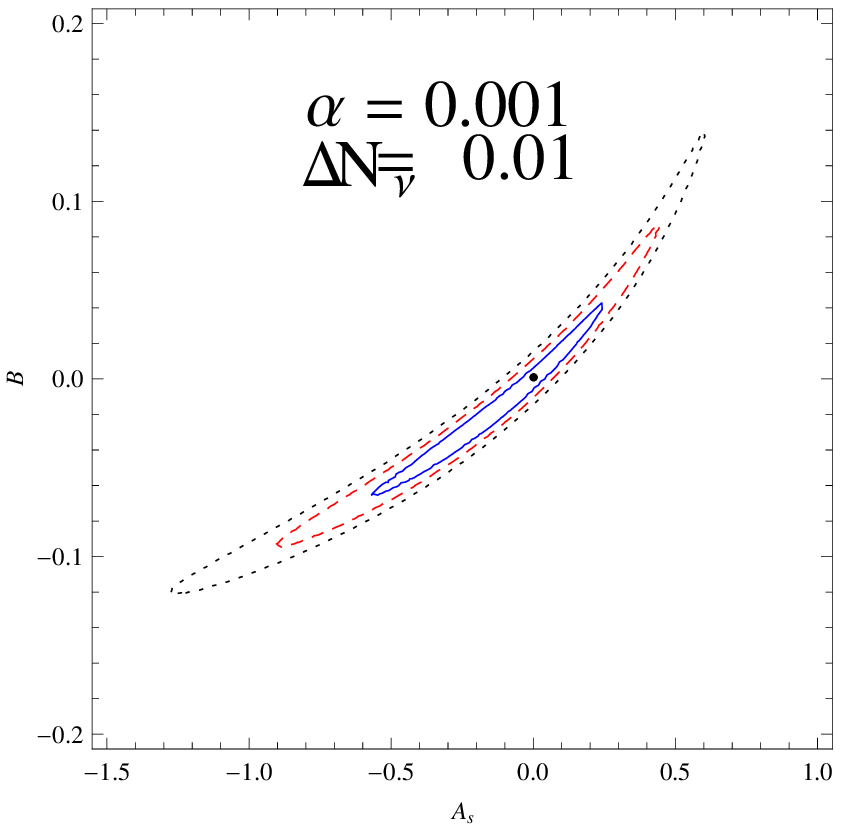}} \\
\subfigure{[b]\includegraphics[width=230pt,height=190pt]{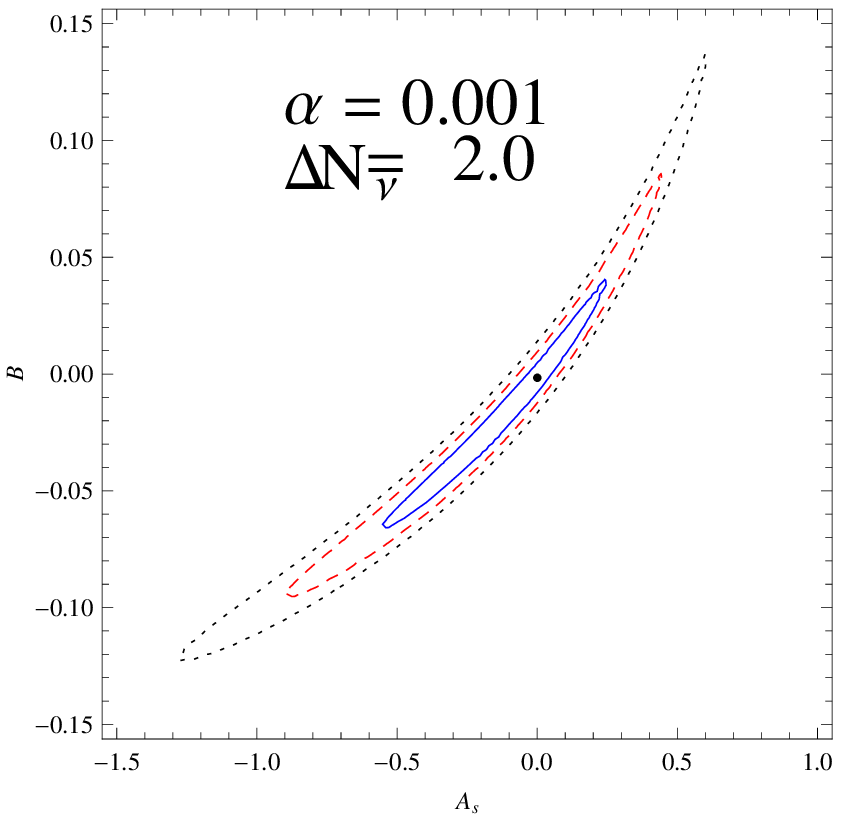}}} \\
\caption{Constraints in beyond-detailed balance for $\alpha=0.001$ from OHD+SDSS+CMB Shift data    $68.3\%$(Solid) $95.4\%$ (Dashed) and $99.73 \%$  (Dotted) contours.}
\label{BDB-d}
\end{figure}

Fig. 7(a) is plotted for $\alpha=0.001$ and $\Delta N_{\nu}$=0.01 for best-fitted value of $\beta$ and $\Omega_{Ko}$,  it is evident that the permissible values of $B$ now lies in the range $-0.06682 <B<0.0407$, $-0.09434 <B<0.08918$, $-0.1206 <B<0.1416$  at $68.3\%$, $95.4\%$, $99.73\%$ confidence levels respectively. 
Fig. 7(b)  is plotted for $\alpha=0.001$ and $\Delta N_{\nu}$=2.0,  it is evident that the values of $B$ lies in the range $-0.06347 <B<0.04844$, $-0.09244 <B<0.09241$, $-0.1194<B<0.1414$  at $68.3\%$, $95.4\%$, $99.73\%$ confidence levels respectively.
The contours drawn in figs 7(a) and 7(b) show that the range of permissible values of $ B $ now decreases with an increase in the  effective neutrino parameter. We note that the allowed range of values of the parameter $B$,  decreased appreciably here compared to that obtained from  figs. 5 (a)-5 (c). This signifies the fact that as the contribution of dark radiation increases (through effective neutrino parameter)  the range of admissible values  of $B$ decreases in the case of  beyond detailed-balance scenario which is same as one obtains in the case of detailed balance scenario.

\section{Viability of MCG in HL gravity}

In this section we discuss some of the implications of the present scenario. Here we determine the evolution of the equation of state parameter of the total cosmic fluid of the universe which is  defined as $w(z)=\frac{p_{tot}}{\rho_{tot}}$, with the total pressure and energy density in the case of detailed-balance scenario. The pressure and energy density  are given by
\begin{equation}
p_{tot}=p_{c}+\frac{1}{3}\rho_{r}+\frac{2}{\kappa^{2}}\left[\frac{K^{2}}{\Lambda a^{4}}-3\Lambda\right],
\end{equation}

\begin{equation}
\rho_{tot}=\rho_{c}+\rho_{b}+\rho_{r}+\frac{2}{\kappa^{2}}\left[\frac{3K^{2}}{\Lambda a^{4}}+3\Lambda\right].
\end{equation}

In the case of beyond-detailed balance scenario the total pressure and the energy density is 
given respectively as
\begin{eqnarray}
p_{tot} &=& p_{c}+\frac{1}{3}\rho_{r}  \nonumber \\ 
 &&+ \left[-\frac{\sigma_{1}}{6\sigma_{0}}+\frac{\sigma_{3}K^{2}}{18\sigma_{0}a^4} +\frac{\sigma_{4}K}{6\sigma_{0}a^6}\right],
\end{eqnarray} 

\begin{equation}
\rho_{tot} &=& \rho_{c}+\rho_{b}+ \rho_{r}+ \left[\frac{\sigma_{1}}{6\sigma_{0}}+\frac{\sigma_{3}K^{2}}{6\sigma_{0}a^4} +\frac{\sigma_{4}K}{6\sigma_{0}a^6}\right].
\end{equation} 
Here we replace the scale factor by redshift parameter, and the expression for density parameter and the Hubble parameter  are expressed in terms of the  redshift parameter in the equation of state. Consequently we get
\begin{equation}
 w(z) =  \frac{p_{tot}}{\rho_{tot}}.
\end{equation}

\begin{figure}
\centering
{\includegraphics[width=230pt,height=190pt]{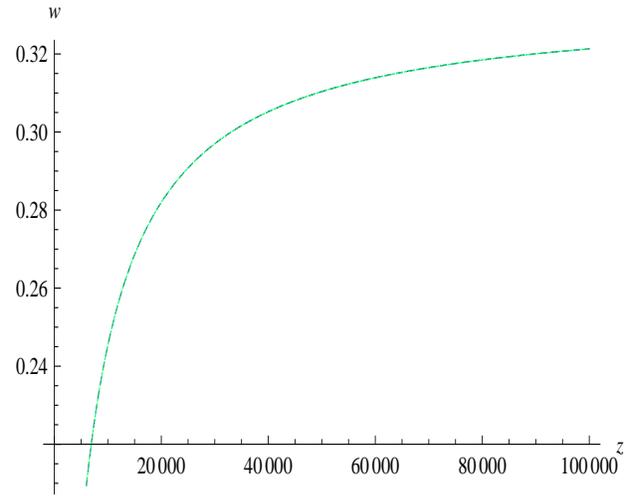} \label{w(z)}}\\
\caption{Equation of state parameter in beyond-detailed balance scenario }.
\label{HL-c}
\end{figure}
From the plot of $w(z)$ with $z$,  for beyond detailed-balance scenario we note that at high redshift 
(i.e., early times) it attains a fixed value $\frac{1}{3}$  since radiation dominates in that epoch. In the intermediate redshift it behaves as dust for quite a long time. It is observed that the equation of state parameter picks up negative values at small redshift, i.e., at very recent past. In the case of closed or open universe the present value of the equastion of state parameter is found to be negative (-0.7) which admits a late accelerating universe.

In order to check the validity of the scenario we employed the best-fit values of the parameters of the MCG to find supernovae magnitudes ($\mu$) at different redshift $(z)$ and plotted $\mu$ vs. $z$ curve. We compared these with original curves of Union2 data  and observed  an excellent agreement.
\begin{figure}
\centering
{\includegraphics[width=230pt,height=190pt]{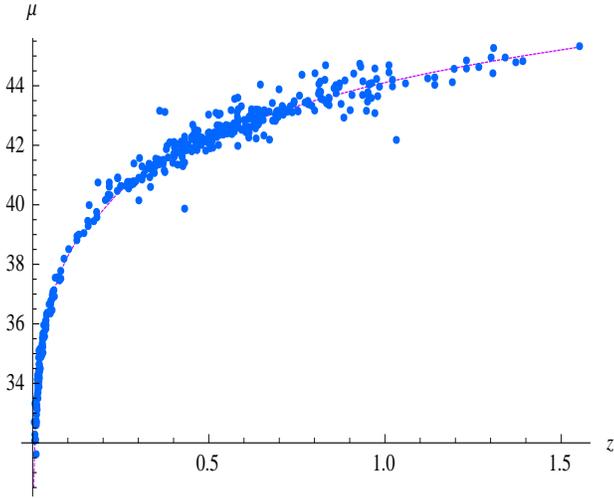} \label{BDB-mu}}\\
\caption{The comparison of the Union2 data with the best-fit values in beyond-detailed balance}
\label{HL-d}
\end{figure}

\section{Discussion}

In this paper we present cosmologies with modified Chaplygin gas (MCG)  in HL gravity scenario taking into account detailed balance and beyond detailed balance conditions both in the presence and absence of dark radiation. The equation of state of MCG has three unknown parameters. The permissible values of these parameters are explored from the observed data. Using data from different observations, namely, $H(z)-z$ (OHD), BAO peak, CMB shift parameter data,  we  determine the admissible values of the EOS parameters considered here. In the MCG we have a parameter $B$ that represents the matter part. We analyze and determine the allowed range of values of $B$ for  viable cosmologies. The analysis is carried out here both in the case of open and closed universe at  $68.3\%$, $95.4\%$, $99.73\%$ confidence levels.

In a close universe we note that the permissible values of $B$ parameter lies in the range $-0.05702 <B<0.0419$, $-0.09988 <B<0.1002$, $-0.1296 <B<0.1595$  at $68.3\%$, $95.4\%$, $99.73\%$ confidence levels respectively for a maximum value of effective neutrino parameter. The range of $B$ obtained here is less than that  obtained for best-fit value of effective neutrino parameter when dark radiation is not taken into account \citep{paulprd}.

It is evident from the figures 2 (a) and 2(b) that the range of $B$  increases with an increase in the effective neutrino parameter $\Delta N_{\nu}$.
In an open universe, the parameter $ B$ lies in the range $-0.05812 <B<0.03751$, $-0.09768 <B<0.09576$, $-0.1274 <B<0.1584$  at $68.3\%$, $95.4\%$, $99.73\%$ confidence levels respectively for maximum value of effective neutrino parameter. The domain of $B$ is found to be less than that  obtained for best-fit value of effective neutrino parameter without dark radiation  \citep{paulprd}.

It appears from the analysis of a  close and an open universe that in a close universe the domain of admissible values $B$ is comparatively narrower than  that of an open universe at  $68.3\%$, $95.4\%$, $99.73\%$ confidence levels respectively. We alaso note that the  $B$ may take  negative values in this case. The negative value of $B$  implies existence of exotic matter. In the  literatures \citep{fabris} the acceptable value of $B$ as was predicted to be very small, which once again gets support from our analysis.

In the beyond-detailed balance scenario there are six free parameters, namely $\Omega_{Ko}$ , $A_{S}$, $B $, $\alpha$, $\beta$, 
$\Delta N_{\nu}$. It is found  that the entire range of effective neutrino parameter is consistent with observations from our numerical analysis. The dependence of the extreme values of the neutrino parameter on other parameters are also  shown in figs. (6) and (7).

The contours drawn in fig. (5) for beyond-detailed balance with different $\alpha$ for best-fitted value of $\beta$, $\Delta N_{\nu}$  and $\Omega_{Ko}$ projects the admissible values of $B$ which lies in the range $-0.09257 <B<0.0707$, $-0.1326 <B<0.1493$, $-0.1727<B<0.2247$  at $68.3\%$, $95.4\%$, $99.73\%$ confidence levels respectively.
Thus the range of  $B$ in this case become larger than that of  detailed balance scenario without dark radiation \citep{paulprd}.

The contours drawn in fig. 6(a) which is plotted for $\alpha=0.999$ and $\Delta N_{\nu}$=0.01 for best-fitted value of $\beta$ and $\Omega_{Ko}$,  projects the admissible values of $B$ which  lies in the range $-0.05338 <B<0.03351$, $-0.07935 <B<0.07445$, $-0.1013 <B<0.1224$  at $68.3\%$, $95.4\%$, $99.73\%$ confidence levels respectively. 
The contours drawn in fig. 6(b) which is plotted for $\alpha=0.999$ and $\Delta N_{\nu}$=2.0, permits the  values of the parameter $B$ which  lies in the range $-0.05462 <B<0.03617$, $-0.07978 <B<0.07662$, $-0.1032<B<0.1162$  at $68.3\%$, $95.4\%$, $99.73 \%$ confidence levels respectively.
It is clear that the range of values of $B$ decreases with an increase of the effective neutrino parameter.

The contours drawn in fig. 7(a) which is plotted for $\alpha=0.001$ and $\Delta N_{\nu}$=0.01 for best-fitted value of $\beta$ and $\Omega_{Ko}$ gives the allowed  values of $B$ which lies in the range $-0.06682 <B<0.0407$, $-0.09434 <B<0.08918$, $-0.1206 <B<0.1416$  at $68.3\%$, $95.4\%$, $99.73\%$ confidence levels respectively. 
The contours drawn in fig. 7(b) which is plotted for $\alpha=0.001$ and $\Delta N_{\nu}$=2.0 gives the allowed  values of $B$ which  lies in the range $-0.06347 <B<0.04844$, $-0.09244 <B<0.09241$, $-0.1194<B<0.1414$  at $68.3\%$, $95.4\%$, $99.73\%$ confidence levels respectively.
It is clearly visible from above analysis that the range of values of $B$ decreases with an increase in effective neutrino parameter. 

We note that the range of values of $B$ decreases appreciably here compared to that obtained from the figs. 5 (a)- 5 (c). This signifies that as the contribution of dark radiation increases (through effective neutrino parameter) the contribution to the permissible range of values of $B$ decreases  in beyond detailed-balance scenario like that one obtains in the case of  detailed balance scenario \citep{paulprd}.
In figure (8) we plot the variation of the total equation of state parameter $w(z)$ with  redshift parameter $z$ for beyond detailed-balance scenario. The curve shows  the evolutionary phases of the universe efficiently. It is evident that at 
high redshift (early times) the of equation of state parameter attains  $\frac{1}{3}$, indicating radiation domination in that epoch. However, in the intermediate redshift we note that dust dominates through MCG for quite a long  period of time.
The best fit values of $B$, $A_{S}$ and $\Delta N_{\nu}$ are shown in tables (1) and (2) for close universe and tables (3) and (4) are shown for open universe.

Using the best-fit values in beyond-detailed balance scenario $\mu$ vs redshift curve is plotted in fig. (9) and the figure is compared  with Union Compilation data. It is evident from the figure that cosmologies in Horava-Lifshitz gravity with  MCG fits well with the   experimental result. 

In this analysis we studied the dependance of extreme values of $\Delta N_{\nu}$ on other parameters in detailed balance scenario with MCG. We also present here results obtained on the beyond-detailed balance scenario. In the case of beyond-detailed balance scenario (BDB) there are two more free parameters and found that the theory is rich and all the features of cosmologies can be accommodated with MCG.  Earlier in the Einstein-frame, MCG is employed to obtain  viable cosmological models \citep{thakur,lu}. Here MCG is employed  in the HL gravity and determined various physical parameters of the universe which gets supports from observations. However, the present analysis does not enlighten the conceptual issues in HL gravity. It is important to look into  details why the neutrino parameter is  very small in HL gravity with MCG which will be discussed elsewhere.

\section{Acknowledgements}
The authors would like to thank {\it IUCAA Reference Centre}  at North Bengal University for extending necessary research facilities to initiate the work. BCP would like to thank {\it Inter-University Centre for Astronomy $\&$ Astrphysics} (IUCAA), Pune for hospitality to complete the work.

\end{document}